\documentstyle[12pt]{article}

\newcommand{\EQ}{\begin{equation}}
\newcommand{\EN}{\end{equation}}
\newcommand{\bea}{\begin{eqnarray}}
\newcommand{\ena}{\end{eqnarray}}
\newcommand{\vs}[1]{\vspace{#1 mm}}

\renewcommand{\b}{\beta}
\renewcommand{\c}{\gamma}

\newcommand{\shalf}{\frac{1}{2}}
\newcommand{\pa}{\partial}
\newcommand{\dz}{\frac{dz}{2\pi i}}
\newcommand{\nn}{\nonumber \\}

\begin{document}

\topmargin 0pt
\oddsidemargin 5mm

\renewcommand{\Im}{{\rm Im}\,}
\newcommand{\NP}[1]{Nucl.\ Phys.\ {\bf #1}}
\newcommand{\PL}[1]{Phys.\ Lett.\ {\bf #1}}
\newcommand{\CMP}[1]{Comm.\ Math.\ Phys.\ {\bf #1}}
\newcommand{\PR}[1]{Phys.\ Rev.\ {\bf #1}}
\newcommand{\PRL}[1]{Phys.\ Rev.\ Lett.\ {\bf #1}}
\newcommand{\PTP}[1]{Prog.\ Theor.\ Phys.\ {\bf #1}}
\newcommand{\PTPS}[1]{Prog.\ Theor.\ Phys.\ Suppl.\ {\bf #1}}
\newcommand{\MPL}[1]{Mod.\ Phys.\ Lett.\ {\bf #1}}
\newcommand{\IJMP}[1]{Int.\ Jour.\ Mod.\ Phys.\ {\bf #1}}

\begin{titlepage}
\setcounter{page}{0}
\begin{flushright}
NBI-HE-93-76\\
hep-th/9312187
\end{flushright}

\vs{15}
\begin{center}
{\Large $N=1$ FROM $N=2$ SUPERSTRINGS}
\vs{15}

{\large Nobuyoshi Ohta\footnote{e-mail address:
ohta@nbivax.nbi.dk}$^,$\footnote{
Permanent address: Institute of Physics, College of General Education,
Osaka University, Toyonaka, Osaka 560, Japan}}\\

\vs{8}
{\em NORDITA, Blegdamsvej 17, DK-2100 Copenhagen \O, Denmark}\\
\vs{8}

{\large Jens Lyng Petersen\footnote{e-mail address:
 jenslyng@nbivax.nbi.dk}}\\
\vs{8}
{\em The Niels Bohr Institute, Blegdamsvej 17, DK-2100 Copenhagen \O,
Denmark}
\end{center}

\vs{8}
\centerline{{\bf{Abstract}}}

We give a simple proof that a particular class of $N=2$ superstrings are
equivalent to the $N=1$ superstrings. This is achieved by constructing a
similarity transformation which transforms the $N=2$ BRST operators into
a direct sum of the BRST operators for the $N=1$ string and topological
sectors.

\end{titlepage}
\newpage
\renewcommand{\thefootnote}{\arabic{footnote}}
\setcounter{footnote}{0}

One of the long standing problems in string theories is how to give
a formalism which allows to interpolate between various string theories.
Recently a very interesting discovery in this direction has been made
\cite{BV}. It has been shown that the $N=0 (N=1)$ strings can be viewed
as a special class of vacua for $N=1 (N=2)$ superstrings. The equivalence
of these $N=1$ and $N=0$ strings has been discussed in the original paper
\cite{BV} and is further confirmed in refs.~\cite{FO,IK} (see also
\cite{FB}).

The purpose of this paper is to give a simple and explicit proof that
the above choice of the vacua in $N=2$ superstrings is equivalent to the
$N=1$ superstring. That this is true has also been argued in ref.~\cite{BV}
for the scattering amplitudes. The argument, however, is rather complicated
involving picture-changing operators and instanton-number-changing operators
and it is not clear if the operator algebras are also isomorphic.

The most straightforward and transparent approach to this problem is the one
in ref.~\cite{IK}, where it has been shown that the BRST operator for the
$N=1$ superstrings is transformed into a direct sum of the BRST operators
for the $N=0$ string and topological sectors by a similarity transformation.
We will show that it is also possible to construct such a similarity
transformation for the $N=2$ superstrings and demonstrate the equivalence.
This may appear straightforward but, since the argument in ref.~\cite{BV}
is involved and it is not clear if that argument is applicable to the
operator algebra, it is better to give a simpler proof at the operator level.
We also find that the proof is not quite trivial.

Let us start by recapitulating the $N=2$ formulation of the $N=1$
superstrings in ref.~\cite{BV}. Take the $N=1$ superstrings with super
stress-energy matter tensors $T_m$ and $G_m$ with central charge $c_m=15$, and
add to the system fermionic fields $(\eta_1,\xi_1)$ of spin $(\frac{3}{2},
-\shalf)$ and bosonic $(\b_1,\c_1)$ of spin $(1,0)$.\footnote{We slightly
change the notation of the fields from ref.~\cite{BV}.} This gives a system
with central charge 6 which can then be used as a matter sector for an $N=2$
superstring. Indeed, it is possible to construct $N=2$ generators for this
system. For this purpose, it is convenient to use the bosonization
\EQ
\c_1 = \eta_2 e^{\phi}, \;\;\;
\b_1 = e^{-\phi} \pa\xi_2.
\EN
The $N=2$ generators of the system are then~\cite{BV}
\bea
G^- &=& \eta_1, \nonumber \\
G^+ &=& \c_1 G_m + \xi_1 \left( T_m - \frac{3}{2}\b_1\pa\c_1
      - \shalf \pa\b_1\c_1 \right) -\c_1^2\eta_1 + \pa (\xi_1\xi_2\eta_2)
      + \pa^2\xi_1 + \eta_1\xi_1\pa\xi_1, \nonumber \\
T &=& T_m - \frac{3}{2}\b_1\pa\c_1 - \shalf \pa\b_1\c_1 - \eta_1\pa\xi_1
      - \shalf\pa (\eta_1\xi_1 + \eta_2\xi_2), \nonumber \\
J &=& - \eta_1\xi_1 + \eta_2\xi_2.
\ena
Here the generators $G^+$ and $T$ are written in mixed notation, and it is
more convenient to express these solely in terms of independent fields
$(\eta_1,\xi_1)$, $(\eta_2,\xi_2)$ and $\phi$:
\bea
G^+ &=& \eta_2 e^\phi G_m + \xi_1 \left( T_m + \pa\xi_2\eta_2
      - \shalf (\pa\phi)^2 -\pa^2\phi \right)\nonumber\\
&&    - \eta_1\eta_2\pa\eta_2 e^{2\phi} + \pa (\xi_1\xi_2\eta_2)
      + \pa^2\xi_1 + \eta_1\xi_1\pa\xi_1, \nonumber \\
T &=& T_m - \frac{3}{2}\eta_1\pa\xi_1 - \shalf \pa\eta_1\xi_1
      - \frac{3}{2}\eta_2\pa\xi_2 - \shalf \pa\eta_2\xi_2
      -\shalf(\pa\phi)^2 -\pa^2\phi.
\label{matterpart}
\ena

The BRST operator for this $N=2$ superstring takes the form
\bea
Q_{N=2} =&& \oint\dz \left[
      c \left( T + \shalf T_{gh} \right)
      + \frac{1}{\sqrt{2}} \c^+ \left( G^- + \shalf G^-_{gh} \right)\right.
       \nonumber \\
&&  + \left. \frac{1}{\sqrt{2}} \c^- \left( G^+ + \shalf G^+_{gh} \right)
      -\shalf c_1 \left( J + \shalf J_{gh} \right) \right],
\ena
where $(b,c),(\b^\pm,\c^\mp)$ and $(b_1,c_1)$ are the reparametrization,
supersymmetry and $U(1)$ ghosts, respectively, with the correlations
\EQ
\c^\pm(z)\b^\mp(w) \sim c_1(z)b_1(w) \sim \frac{1}{z-w},
\EN
and the generators with subscript $gh$ are those for ghosts:
\bea
T_{gh} &=& -2b\pa c -\pa b c -\frac{3}{2}\b^+\pa\c^- -\shalf\pa\b^+\c^-
 \nonumber\\
 && -\frac{3}{2}\b^-\pa\c^+ -\shalf\pa\b^-\c^+ -b_1\pa c_1, \nonumber\\
G^{\pm}_{gh} &=& \frac{1}{\sqrt{2}}(\mp \b^\pm c_1 \mp \pa b_1\c^\pm
  \mp 2b_1\pa\c^\pm + 3\b^\pm\pa c +2\pa\b^\pm c -b\c^\pm), \nonumber\\
J_{gh} &=& \b^+\c^- -\b^-\c^+ +2\pa(b_1 c).
\label{ghostpart}
\ena
Substituting (2), (3) and (6) into (4), we find
\bea
Q_{N=2} =&& \oint \dz \left[ \left( c +\frac{1}{\sqrt{2}}\c^-\xi_1 \right)
 T_m  +\frac{1}{\sqrt{2}}\c^- \eta_2 e^\phi G_m +bc\pa c -\shalf b\c^+\c^-
 \right.  \nonumber\\
&& -c(\b^+\pa\c^- + \b^-\pa\c^+) +\shalf\pa c(\b^+\c^- + \b^-\c^+) \nonumber\\
&& + c\left(-\frac{3}{2}\eta_1\pa\xi_1 - \shalf \pa\eta_1\xi_1
   - \frac{3}{2}\eta_2\pa\xi_2 - \shalf \pa\eta_2\xi_2
   - \shalf(\pa\phi)^2 -\pa^2\phi \right) \nonumber \\
&& +\frac{1}{\sqrt{2}}\c^-\left[ \xi_1 \pa\xi_2\eta_2 + \pa(\xi_1\xi_2\eta_2)
   - \xi_1\left(\shalf (\pa\phi)^2 +\pa^2\phi \right)
   - \eta_1\eta_2\pa\eta_2 e^{2\phi} \right.\nonumber\\
&& \left. + \pa^2\xi_1 + \eta_1\xi_1\pa\xi_1 \right]
   + \frac{1}{\sqrt{2}}\c^+\eta_1 + \shalf c_1(\eta_1\xi_1 -\eta_2\xi_2
   +\c^+\b^- -\c^-\b^+) \nonumber\\
&& \left. +c\pa c_1 b_1 +\shalf b_1(\c^+\pa\c^--\pa\c^+\c^-)
\right].
\ena

We would like to compare this with the BRST operator for the $N=1$ superstring:
\EQ
Q_{N=1} = \oint\dz \left[
cT_m -\shalf\c G_m +bc\pa c -\frac{1}{4}b\c^2 +\shalf \pa c\b\c -c\b\pa\c
\right].
\EN
Comparison between (7) and (8) suggests the identification
\EQ
\c = -\sqrt{2}\c^-\eta_2 e^\phi, \;\;\;
\b = -\frac{1}{\sqrt{2}} e^{-\phi} \xi_2\b^+.
\EN
The $\b$ in this relation is chosen so as to give the correct operator
product with $\c$. Substituting (9) into (8), we can express the $N=1$ BRST
operator (8) in terms of the $N=2$ fields:
\bea
Q_{N=1} =&& \oint \dz \left[ c T_m +\frac{1}{\sqrt{2}}\c^- \eta_2 e^\phi G_m
  +bc\pa c - c\b^+\pa\c^- +\shalf\pa c\b^+\c^- \right. \nonumber\\
&& + c\left(-\frac{3}{2}\pa\eta_2\xi_2 - \shalf \eta_2\pa\xi_2
  - \shalf(\pa\phi)^2 -\pa^2\phi \right) \nonumber\\
&& \left. - \frac{1}{2}b(\c^-)^2\eta_2\pa\eta_2 e^{2\phi}
  - c(\b^+\c^- + \eta_2\xi_2)(\pa\phi + \eta_2\xi_2) \right].
\ena

Now our claim is that the BRST operator (7) is mapped into a direct sum
of those for the $N=1$ superstrings (10) and topological sectors by a
similarity transformation
\EQ
e^{R} Q_{N=2} e^{-R} = Q_{N=1}+Q_{top}+Q_{U(1)},
\EN
where
\bea
R &=& \oint\dz \left[ \frac{1}{\sqrt{2}}\xi_1
  (-\c^- b+ 3\pa c\b^- + 2c\pa\b^- +\b^-c_1+2b_1\pa\c^- +\pa b_1\c^-)
  \right. \nonumber \\
&& - \left. \left( 2b_1c + \frac{1}{\sqrt{2}}\b^-c\xi_1
  +\frac{1}{\sqrt{2}}\c^-b_1\xi_1 \right)(\pa\phi +\eta_2\xi_2)
  -\b^-\c^-\eta_2\pa\eta_2 e^{2\phi}\right],
\ena
and
\bea
Q_{top} &=& \oint\dz \frac{1}{\sqrt{2}}\c^+\eta_1,\nonumber \\
Q_{U(1)} &=& -\oint\dz \shalf c_1(\eta_2\xi_2 +\b^+\c^-).
\ena
are the BRST operators for the topological sectors. It is easy to confirm
that the BRST operators on the right hand side of eq. (11) all anticommute
with one another and are nilpotent. With this form of the BRST operator,
it is obvious that the cohomology of the $Q_{N=2}$ is a direct product of
those of $Q_{N=1}, Q_{top}$ and $Q_{U(1)}$. The BRST operator $Q_{top}$
imposes the condition that $\b^-, \c^+, \eta_1$ and $\xi_1$ all decouple
and its cohomology consists of their vacuum alone. The modes $\b^+,
\c^-,\eta_2,\xi_2$ and $\phi$ can be represented in terms of $\b,\c$ in
eq.~(9) as well as $j\equiv (\eta_2\xi_2 +\pa\phi)$ and
$\tilde{j}\equiv (\eta_2\xi_2 +\b^+\c^-)$.
\footnote{The two fermionic degrees of freedom
represented by $\eta_2,\xi_2$ count like a single bosonic degree
of freedom. Also note that the currents, $j$ and $\tilde{j}$, commute with
$\b$ and $\c$ in (9) and represent modes independent of $\b$ and $\c$.}
The latter two together with $b_1$ and $c_1$ decouple from the physical
subspace due to the condition imposed by the BRST operator $Q_{U(1)}$
in (13). Thus we are left with only the particular
combinations of the fields in (9) as well as the fields in the $N=1$
superstrings and obtain one-to-one correspondence between
the cohomologies of $Q_{N=2}$ and $Q_{N=1}$. It is also clear that
the transformation manifestly keeps the operator algebra. This establishes
the equivalence of the $N=1$ superstrings and the $N=2$ superstrings.

It is instructive to rewrite the total energy-momentum tensor for the system.
First, it may be shown that
$$T_{\mbox{tot}}\equiv T+T_{gh}$$
(where $T$ and $T_{gh}$ are given by (\ref{matterpart}) and (\ref{ghostpart}))
in fact remains invariant under the similarity transformation. Second, it is
intersting that it may be written as
\EQ
T_{\mbox{tot}}=T_{N=1}+\{Q_{top},b_{top}\}+\{Q_{U(1)},b_{U(1)}\}
\EN
where
\bea
T_{N=1}&=&T_m-2b\partial c-\partial b c-\frac{3}{2} \beta\partial\gamma-
\frac{1}{2}\partial\beta\gamma\nn
&=& T_m-2b\partial c-\partial b c-\frac{3}{2} \beta^+\partial\gamma^- -
\frac{1}{2}\partial\beta^+\gamma^-\nn
&&-\frac{1}{2}\eta_2\partial\xi_2-\frac{3}{2}\partial\eta_2\xi_2-
\frac{1}{2}(\partial\phi)^2 -\partial^2\phi -\beta^+\gamma^-\partial\phi\nn
&&-\eta_2\xi_2(\partial\phi+\beta^+\gamma^-)\nn
b_{top}&\equiv&-\frac{3}{\sqrt{2}}\beta^-\partial\xi_1-\frac{1}{\sqrt{2}}
\partial\beta^-\xi_1\nn
\{Q_{top},b_{top}\}&=& -\frac{3}{2}\eta_1\partial\xi_1-\frac{1}{2}\partial
\eta_1\xi_1 -\frac{3}{2}\beta^-\partial\gamma^+-\frac{1}{2}\partial\beta^-
\gamma^+\nn
b_{U(1)}&\equiv&-2b_1 j\nn
\{Q_{U(1)},b_{U(1)}\}&=&-b_1\partial c_1+j\tilde{j}
\ena
Here one may work out that
\EQ
j\tilde{j}=\partial\eta_2\xi_2-\eta_2\partial\xi_2+\beta^+\gamma^-
\partial\phi+\eta_2\xi_2(\partial\phi+\beta^+\gamma^-)
\EN
Thus the total energy-momentum tensor for the $N=2$ system is equivalent
to that for the $N=1$ system up to topological terms.

Continuing this line of investigation, it is natural to consider similar
embeddings of strings into $N=4$ superstrings and also into $W$ strings.
An embedding of bosonic string into $W_3$ string has been found in
ref.~\cite{BFW}. We hope to discuss these embeddings
elsewhere.

\newpage
\noindent
{\it Acknowledgements}

We would like to thank F. Bastianelli for valuable discussions and comments,
O. T\"ornkvist for assistance in computation and M. Kato for correspondence.
One of us (N. O.) thanks Paolo Di Vecchia for discussions, support and kind
hospitality at NORDITA where this work was done. The calculations in the
transformation (11) have been performed by the package
{\bf ope.math} for the operator product expansion developed by A. Fujitsu,
whom we thank for sending us his package.
In addition the calculations have been checked using the OPE package
developed by Kris Thielemans, whose software is gratefully acknowledged.


\newpage
\begin{thebibliography}{99}
\bibitem{BV} N. Berkovits and C. Vafa, preprint HUTP-93/A031, KCL-TH-93-13,
      hep-th/9310170 (1993).
\bibitem{FO} J. M. Figuerao-O'Farrill, preprints QMW-PH-93-29, hep-th/9310200
      and QMW-PH-93-30, hep-th/9312033 (1993).
\bibitem{IK} H. Ishikawa and M. Kato, preprint UT-Komaba/93-23, hep-th/9311139
      (1993).
\bibitem{FB} F. Bastianelli, preprint NBI-HE-93-69, hep-th/9311157 (1993).
\bibitem{BFW}N. Berkovits, M. Freeman and P. West, preprint KCL-TH-93-15,
      hep-th/9312013 (1993).
\end{thebibliography}
\end{document}